\documentclass[a4paper,11pt]{article}

\usepackage{jinstpub} 
\usepackage{siunitx}
\usepackage[version=3]{mhchem}
\usepackage{subfig}
\usepackage{systeme}

\title{\boldmath Measure of negative ion density in a large negative ion source using Langmuir probes}

\author[a]{C. Poggi,\note{Corresponding author.}}
\author[a]{M. Spolaore,}
\author[a]{M. Barbisan,}
\author[a]{M. Brombin,}
\author[a]{R. Cavazzana,}
\author[a,b]{N. Marconato,}
\author[a]{R. Pasqualotto,}
\author[a]{A. Pimazzoni,}
\author[a,b]{E. Sartori,}
\author[a]{and G. Serianni}

\affiliation[a]{Consorzio RFX (CNR, ENEA, INFN, Università di Padova, Acciaierie Venete SpA),\\Corso Sati Uniti 4, 35127, Padova, Italy}
\affiliation[b]{Università degli Studi di Padova,\\Via VIII Febbraio 2, 35122, Padova, Italy}

\emailAdd{carlo.poggi@igi.cnr.it}

\abstract{
Neutral Beam Injectors (NBIs) based on negative ions are the workhorses of future fusion reactors, such as ITER, which they are expected to provide with up to \SI{33}{MW} of power to heat the fusion plasma. The negative hydrogen ions are extracted from a RF plasma, in which a magnetic filter field cools down the electrons reaching the so-called expansion region and allows the formation of negative ions near the apertures in the plasma grid. To further improve the production of negative ions, cesium is usually evaporated inside the source and deposited onto the plasma walls, reducing the work function of the surfaces. This dramatically increases the density of negative hydrogen ions near the surfaces, causing the transition to an electronegative plasma in the vicinity of the plasma grid. This condition can be observed with Langmuir probes, which can then be used to provide a local meaurement of negative ion density in the ion source.

In this paper we use the measurements provided by the Langmuir probe sensors embedded in the plasma grid of SPIDER, the prototype ion source of ITER NBIs, to determine the density of negative ions. A fitting method based on the determination of the collection area of the different plasma species is proposed and adapted to SPIDER experimental condition, taking into account the shape of the probes and the local topology of the magnetic field. The method is then applied to the experimental data, determining the densities of the positive and negative ions and of the electrons during a plasma pulse. Finally, a vertical array of four probes in the plasma grid is used to assess the vertical profile of plasma parameters.

}


\proceeding{8$^{\text{th}}$ Symposium on Negative Ion Beams and Sources\\
  October 3-7, 2022\\
  Padova (Italy)}

\begin{document}
\maketitle
\flushbottom

\section{Introduction}
\label{sec:intro}
Neutral beam injectors for future fusion reactors like ITER require charged beams up to \SI{1}{MeV} of energy to be used as precursor~\cite{art:hemsworth_2017}. For this reason they rely on negative ion sources to produce the required extracted beam current density, that for ITER is set at \SI{355}{A/m^2} for neutral hydrogen beams. The SPIDER experiment (Source for the Production of Ions of Deuterium Extracted from a Radio-frequency plasma), hosted at Consorzio RFX in Padova, is the full-scale prototype of ITER neutral beam injectors~\cite{art:toigo_2017,art:sartori_2022}. It consists of a plasma chamber made of 8 cylindrical drivers inside which an inductively coupled plasma discharge is generated using four \SI{1}{MHz} RF generators capable to deliver up to \SI{800}{kW} of total power. The plasma then protrudes inside the so-called expansion chamber towards the plasma grid, from which negative ions are extracted through 1280 apertures, organized in 4 vertical segments of 4 beamlet groups of 80 beamlets each. A mostly horizontal magnetic filter field is generated inside the expansion region by a current flowing vertically in the plasma grid. This magnetic field cools down the electrons generated inside the drivers, allowing the formation and survival of the negative hydrogen ions near the plasma grid apertures. A further electrode, called bias plate, is placed at \SI{1}{cm} from the plasma grid and frames all the beamlet groups. Both the plasma grid and the bias plate electrode can be biased with respect to the other source walls. The control is usually performed over the total current flowing between the electrode and a protection resistor of $\SI{0.6}{\Omega}$~\cite{art:zamengo_2021}. To increase the surface production of negative ions, cesium is evaporated inside the source from three cesium ovens, placed between the plasma grid segments, and it is deposited over the plasma facing surfaces~\cite{art:rizzolo_2019,art:fadone_2022,art:sartori_2022}. Investigating negative ion production in a large negative ion source for fusion is a key aspect for improving source operation. In particular, negative ion density is usually measured using lasers with Cavity Ring-down spectroscopy or from optical measurements, with the drawback of obtaining only measurements integrated along one line of sight. A localized measurement of negative ion density can provide useful information to assess the uniformity of such a large negative ion source.

SPIDER is equipped with several Langmuir probes embedded in the plasma grid and bias plate electrodes~\cite{art:spolaore_2010,art:brombin_2014}. The probes were operated during SPIDER operation, providing information about plasma parameters and source uniformity~\cite{art:poggi_2021}. Some of them were polarized with voltage ramps, in order to measure the current-voltage characteristics. Cesium injection leads to the occurrence of a nearly ion-ion plasma condition near the plasma grid, characterized by large values of the electronegativity parameter at the sheath edge $\alpha=n_-/n_e$, given by the ratio of the negative ion and electron densities. This transition to an ion-ion plasma produces a decrease of the electron saturation branch of the measured Langmuir probe characteristics, so that standard methods to analyze the I-V curves like 4-parameter fitting of the positive ion branch or second derivative analysis of the electron branch cannot be used. In this paper we apply the method described in the work by Bredin et al~\cite{art:bredin_2014} to fit the entire characteristic, taking into account the different magnetizations of electrons and ions which affect the collection area of the different species. This model allows to estimate the density of the various plasma species, including the negative ions, thus providing a localized measurement of negative ion density inside the ion source. The method is described in section~\ref{sec:model}, while its application to SPIDER experimental data is reported in section~\ref{sec:data}.

\section{Model of the Langmuir probe characteristic}
\label{sec:model}
The Langmuir probe I-V characteristic is modelled considering the contributions of the different plasma species: electrons, positive ions and negative ions. Given the densities of positive and negative ions and of the electrons at the sheath edge ($n_+^s$, $n_-^s$ and $n_e^s$ respectively), their masses ($m_+$, $m_-$, $m_e$) and their temperatures ($T_+$, $T_-$, $T_e$), imposing quasi-neutrality ($n_+^s\approx n_-^s+n_e^s$), below the sheath potential $V_s$ the collected electron, positive and negative ion currents behave as:
\begin{subequations}\label{eq:below}
\begin{align}
\label{eq:below:ele}
I_e(V) & = \frac{1}{4}eS_{eff}^ev_e\frac{n_+^s}{1+\alpha_s}\exp\bigg({e\frac{V-V_s}{T_e}}\bigg) & \text{ with } & v_e=\sqrt{\frac{8T_e}{\pi m_e}}
\\
\label{eq:below:pos}
I_+(V) & = eS_{eff}^+(V)u_B^+n_+^s & \text{ with } & u_B^+=\sqrt{\frac{T_e}{m_+}}\sqrt{\frac{1+\alpha_s}{1+\gamma\alpha_s}}
\\
\label{eq:below:neg}
I_-(V) & = eS_{eff}^-(V_s)n_+^su_B^-\frac{\alpha_s}{1+\alpha_s}\exp\bigg({e\frac{V-V_s}{T_-}}\bigg) & \text{ with } & u_B^-=\sqrt{\frac{T_+}{m_-}}
\end{align}
\end{subequations}
where $\alpha_s=n_-^s/n_e^s$ is the electronegativity parameter at the sheath edge, $S_{eff}^e$, $S_{eff}^+$ and $S_{eff}^-$ are the electron, positive and negative ion effective collection areas of the probe, $v_e$ is the mean velocity of a Maxwellian electron population, $u_B^+$ is the modified Bohm velocity considering negative ions and considering also the electron to ion temperature ratio $\gamma=T_e/T_-$, and $u_B^-$ is the negative ion Bohm velocity~\cite{book:chabert_book}. For $V>V_s$ the currents are instead given by:
\begin{subequations}\label{eq:above}
\begin{align}
\label{eq:above:ele}
I_e(V) & = \frac{1}{4}eS_{eff}^ev_e\frac{n_+^s}{1+\alpha_s}\bigg(2\sqrt{e\frac{V-V_s}{\pi T_e}}+\exp{\bigg(e\frac{V-V_s}{T_e}}\bigg)\text{erfc}\bigg({\sqrt{e\frac{V-V_s}{T_e}}}\bigg)\bigg)
\\
\label{eq:above:pos}
I_+(V) & = eS_{eff}^+(V)u_B^+n_+^s\exp\bigg({e\frac{V_s-V}{T_+}}\bigg)
\\
\label{eq:above:neg}
I_-(V) & = eS_{eff}^-(V)n_+^su_B^-\frac{\alpha_s}{1+\alpha_s}
\end{align}
\end{subequations}
with the electron current being described by the standard OML theory~\cite{book:lieberman_book}.

\begin{figure}
\centering 
\subfloat[\label{fig:model_example}]{\includegraphics[width=.45\textwidth]{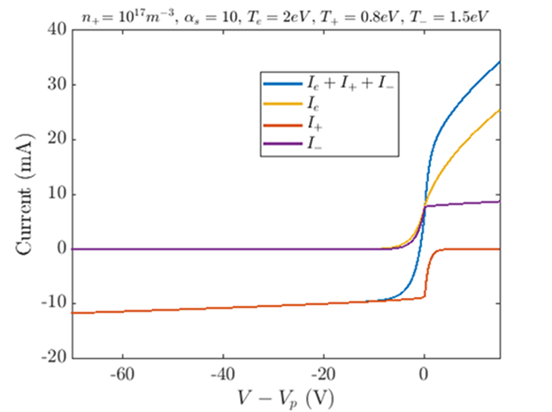}}\qquad
\subfloat[\label{fig:ratio_various}]{\includegraphics[width=.45\textwidth]{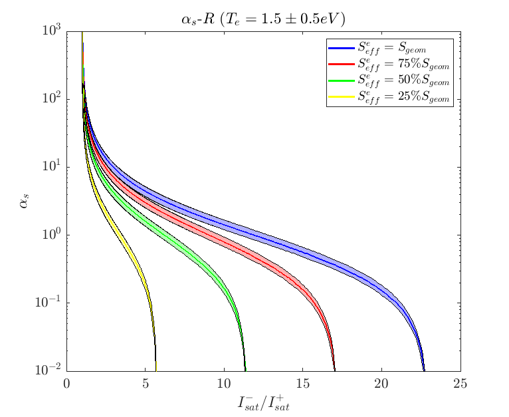}}
\caption{\label{fig:model} (a) Plot of the model described by equations~\ref{eq:below} and \ref{eq:above}, showing the contributions coming from the different plasma species. (b) Dependence of the electronegativity on the ratio between the negative and positive saturation currents, for various values of $S_{eff}^e$.}
\end{figure}

An example showing the various contributions of equations~\ref{eq:below} and \ref{eq:above} to the total current is presented in figure~\ref{fig:model_example} (parameters are $n_+^s=\SI{1e17}{m^{-3}}$, $\alpha_s=\si{10}$, $T_e=\SI{2}{eV}$, $T_+=\SI{0.8}{eV}$, $T_-=\SI{1.5}{eV}$, $m_+=\SI{1.8}{amu}$). Fitting a voltage-current characteristic with this model can in principle provide sheath density and temperature for all the various species that compose the plasma. However, it is necessary to carefully estimate the effective collection areas of all the species, as it is evident from figure~\ref{fig:ratio_various}: here the electronegativity is plotted against the ratio between the negative and positive saturation currents (the temperatures are $T_e=\si{2}\pm\SI{ 0.5}{eV}$, $T_+=\SI{0.8}{eV}$, $T_-=\SI{1.5}{eV}$, $T_-=\SI{1.5}{eV}$) for different values of $S_{eff}^e/S_{geom}$, showing a great variation of the trends for an electronegativity ranging from 0 to 10. Next section is dedicated to describing the calculation of the effective areas for all the species.

\subsection{Evaluation of the collection areas}
The Langmuir probe electrodes embedded in SPIDER plasma grid and bias plate consist of cylinders with a radius $R_{pr}=\SI{3.5}{mm}$ protruding by an height $h=\SI{1}{mm}$ inside the plasma, for a surface exposed to the plasma $S_{geom}=\SI{60}{mm^2}$. Each probe is embedded and isolated within the corresponding plasma facing grid with a gap of \SI{0.25}{mm} surrounding the electrode.

\begin{figure}
\centering 
\subfloat[\label{fig:probessheath_scheme}]{\includegraphics[width=.45\textwidth]{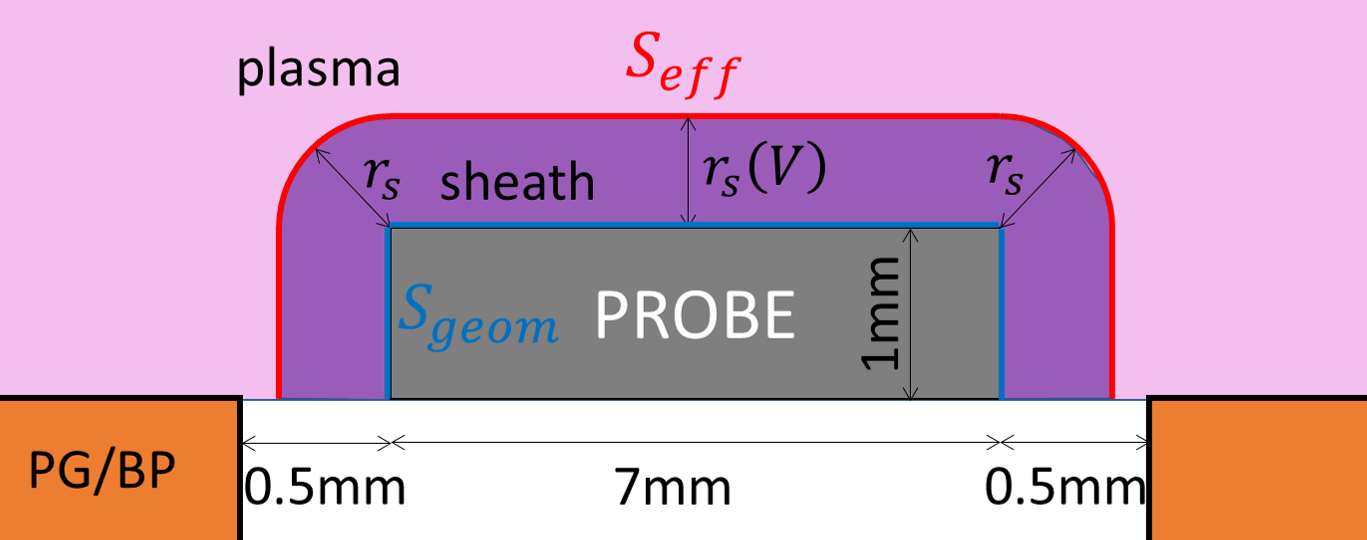}}\qquad
\subfloat[\label{fig:probessheath_sheath}]{\includegraphics[width=.45\textwidth]{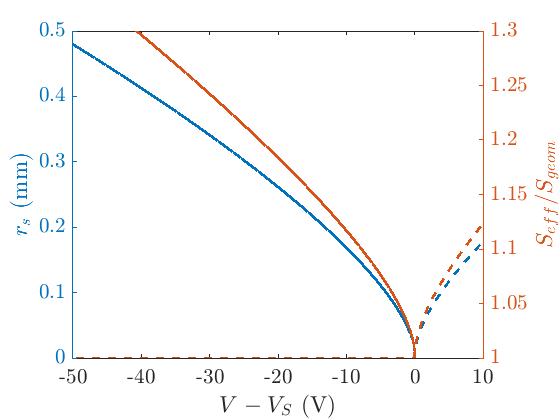}}
\caption{\label{fig:probesheath} (a) Schematic section of a SPIDER Langmuir probe, showing its dimensions and the difference between the geometrical and the effective ion collection areas. (b) Dependence of the sheath size and of the $S_{eff}^{\pm}/S_{geom}$ ratio on the probe polarization voltage.}
\end{figure}

For the positive ions, the collection area is given by the geometrical surface for probe polarization at the plasma potential or above, while for $V<V_s$ it increases with the positive sheath size $r_s^+$, as represented in figure~\ref{fig:probessheath_scheme}. The increase can be calculated as
\begin{equation}
\label{eq:area_increase}
S_{eff}^+=S_{geom}+2\pi r_s(V)h+\pi^2R_{pr}r_s(V)+2\pi r_s^{2}(V)
\end{equation}
that accounts for the increase in the lateral area of the cylinder ($2\pi r_s(V)h$) and for the rounded area of curvature radius $r_s$ (the term $\pi^2R_{pr}r_s(V)+2\pi r_s^2(V)$ corresponds to the area of a quarter of torus).

The $r_s^+$ is calculated solving a 1-dimensional Child-Langmuir model~\cite{art:bredin_2014}:
\begin{equation}\label{eq:child}
\syslineskipcoeff{2}\systeme{
\dfrac{dV}{dx}=-E,
\dfrac{du}{dx}=\dfrac{eE}{m_+u(x)},
\dfrac{dE}{dx}=\dfrac{n_+^seu_B^+}{\epsilon_0u(x)}
}
\end{equation}
with starting conditions $u(0)=u_B^+$ and $E(0)=0$. The sheath size is then the $x$ coordinate corresponding to $V-V_s$.

A similar approach is used to calculate the sheath size for the negative ions, solving the system in \eqref{eq:child} for $V>V_s$ with the negative ion Bohm velocity $u_B^-$ and the negative ion sheath density $n_-^s$, and setting $r_s^-=0$ for $V<V_s$. The plots of the sheath sizes $r_s^\pm$ and the corresponding $S_{eff}^{\pm}/S_{geom}$ are shown in figure~\ref{fig:probessheath_sheath}, covering the typical range of the probe polarization voltage for a plasma with $n_+^s=\SI{1e17}{m^{-3}}$, $\alpha_s=5$, $T_e=\SI{2}{eV}$, $T_+=\SI{0.8}{eV}$, $T_-=\SI{1.5}{eV}$ and an effective positive ion mass $m_+=\SI{1.8}{a.m.u.}$. The sheath size can increase the collection area by more than \SI{30}{\%} at large polarization values as shown in the graph and therefore needs to be taken into account. Furthermore, the sheath increase can in principle be different for positive and negative ions.

A different approach is used to estimate the electron effective collection area $S_{eff}^e$. The magnetic field in correspondence of the probes position is given by the superposition of the filter field generated by the plasma grid current, and the co-extracted electron suppression magnets embedded in the extraction grid, and it presents a significant variation over the probe size, as it is shown in Figure (a). The total magnetic field ranges between 2 and \SI{8}{mT} over the probe surface, while the electron temperature is around $T_e=\SI{2}{eV}$. In this condition, the electron cyclotron frequency $f_{ce}$ is between 56 and \SI{230}{MHz}, and the average gyroradius $R_{ce}$ is between 2.7 and \SI{0.7}{mm} for the particles with speed $\sqrt{kT_e/m_e}$. This value is smaller than the \SI{3.5}{mm} radius of the probe, but larger than its \SI{1}{mm} height, and it is therefore not straightforward to determine if the electrons are magnetized or not with respect to the probe size: which is the most relevant size to consider depends on the direction of the magnetic field lines intercepting the probe surface, and this depends on the probe location and the intensity of the filter field current.

\begin{figure}
\centering 
\subfloat[\label{fig:BatPG}]{\includegraphics[width=.45\textwidth]{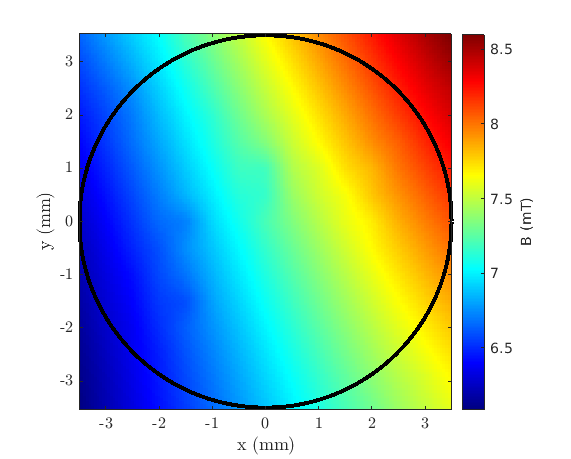}}\qquad
\subfloat[\label{fig:BcompPG}]{\includegraphics[width=.45\textwidth]{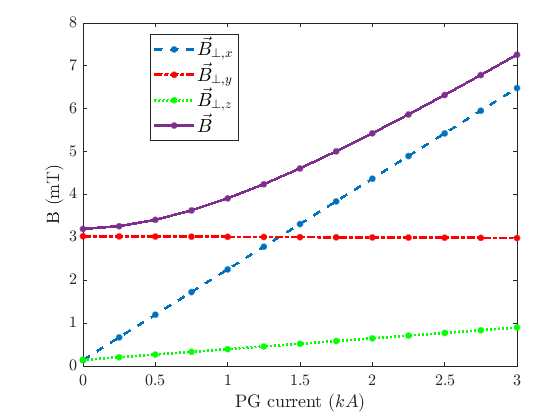}}
\caption{\label{fig:Batprobes} (a) Magnitude of the magnetic field on the surface of a plasma grid probe, for $I_{PG}=\SI{3}{kA}$. (b) Cartesian components and magnitude of the average magnetic field perpendicular to the probe, $\Vec{B}_\perp$, for different values of plasma grid current.}
\end{figure}

The intensity of the magnetic field was calculated numerically using COMSOL and integrated over the cylindrical surface of the probe in order to determine the average magnetic field on the probe surface 
\begin{equation}
\label{eq:Bperp}
    \Vec{B}_\perp=\int_{S_{geom}}\frac{\Vec{B}}{S_{geom}}dS
\end{equation}
and the surface perpendicular to the magnetic field 
\begin{equation}
\label{eq:Sperp}
    S_\perp=\int_{S_{geom}}\frac{|\Vec{B}\cdot\hat{n}|}{B}dS
\end{equation}
with $\hat{n}$ the normal to the surface. From this, an effective probe dimension can be determined as $R_{eff}=\sqrt{S_\perp/\pi}$, to be compared with the electron gyroradius calculated for the perpendicular magnetic field $B_\perp$. Using numerical simulations of the magnetic field~\cite{art:marconato_2022} allows to take into account the correct topology of the magnetic field in SPIDER expansion region and in particular at probe location: due to the superposition of the contributions coming from the permanent magnets embedded in the extraction grid and the current flowing in the plasma grid, the magnetic field intensity and direction can significantly vary even over the probe size, as it is shown in Figure \ref{fig:BatPG}, where the intensity of the magnetic field over the top surface of one of the plasma grid probe in SPIDER bottom segment is presented. The cartesian components of the $\Vec{B}_\perp$ field for various $I_{PG}$ are instead plotted in figure~\ref{fig:BcompPG}. While the PG current mostly modifies the horizontal $x$ component of the field, the others have a comparable magnitude, with the total field spanning from 3 to \SI{7}{mT}. The other plasma grid probes present similar values of magnetic field with small differences due to the exact probe location.

According to Usoltceva et al~\cite{art:usoltceva_2018} the factor $\beta=R_{eff}/R_{ce}$ controls the passing of $S_{eff}^e$ from $S_{geom}$ to $S_\perp$ as effective surface for the electron collection at plasma potential, following the equation
\begin{equation}
    \label{eq:usolt}
    S_{eff}^e=S_{geom}\exp(-\beta^2/2)+S_\perp(1-\exp(-\beta^2/2)).
\end{equation}

\begin{figure}
\centering 
\subfloat[\label{fig:expbeta}]{\includegraphics[width=.45\textwidth]{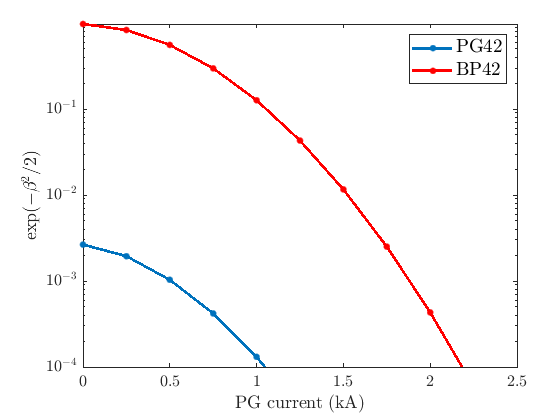}}\qquad
\subfloat[\label{fig:sperp}]{\includegraphics[width=.45\textwidth]{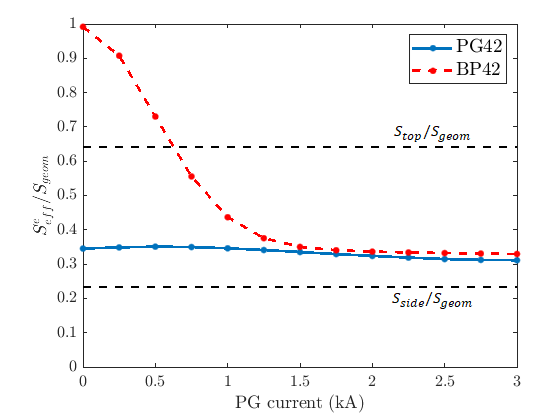}}
\caption{\label{fig:beta_sperp} Dependence on $I_{PG}$ of (a) $\exp(-\beta^2/2)$ assuming $T_e=\SI{2}{eV}$ and (b) of the ratio $S_\perp/S_{geom}$ for a plasma grid and a bias plate probe located in the bottom segment of SPIDER. }
\end{figure}

The same approach is applied in this paper, using the simulations of the magnetic field over the probe surface to estimate both $R_{eff}$ and $S_\perp$. As it can be noticed from figure~\ref{fig:BcompPG}, the average magnetic field can have different orientations with respect to the probe axis. A typical trend of $\exp(-\beta^2/2)$ as a function of $I_{PG}$ for a probe embedded in the plasma grid (PG42) and another embedded in the bias plate (BP42, at approximately \SI{10}{mm} from PG42 along z direction), with an electron temperature of \SI{2}{eV} is shown in figure~\ref{fig:expbeta}, while figure~\ref{fig:sperp} presents the corresponding $S_{eff}^e/S_{geom}$ ratio. As it can be noticed, while for the plasma grid probe it is always below $1\%$, for the bias plate probe it can go up to 1 for low plasma grid currents, and goes below $10\%$ for $I_{PG}>1kA$. The black dashed lines show the fraction of the total surface that corresponds to the top surface of the probe (top line) and to the projection of the lateral surface (bottom line). As it can be noticed, for both probes the effective area calculated with equation~\ref{eq:usolt} is different from those reference values.

With this approach, it was possible to determine the effective collecting areas of the different plasma species, starting from the measurement of the plasma grid current and assuming the effective mass of the positive ions to be known.

\subsection{Fitting of the data and error estimation}
The model described in the previous section was applied to the Langmuir probe data acquired during SPIDER campaigns. A non-linear least-square algorithm was used to estimate the fitting parameters $n_+^s$, $\alpha_s$, $T_e$ and $V_s$ for a measured voltage-current characteristic. The positive and negative ion temperature were fixed at \si{0.8} and \SI{1.5}{eV} respectively, as they cannot be easily derived from the Langmuir probe data (although they are present in the model described in previous section). The effective positive ion mass is set at $m_+=\SI{1.8}{a.m.u.}$ as usually done for similar ion sources \cite{art:schiesko_2012}. However, this quantity was never directly measured in SPIDER. As it can in principle vary between 1 and \SI{3}{a.m.u.}, depending on the relative fraction of $H^+$, $H_2^+$ and $H_3^+$, a \SI{20}{\%} RMS error bar was assumed for it. Concerning the effective electron collection area described above, this depends on the accuracy of the magnetic field simulations, which was taken into account by assuming a \SI{10}{\%} RMS error on the value of $S_\perp$. To assess the effect of these error sources to the parameter estimation, fits for randomly generated values of $m_+$ and $S_\perp$ were performed on selected voltage-current characteristics, and the RMS of the corresponding resulting parameters distributions were calculated. This yielded \SI{0.1}{eV} of RMS error on $T_e$, a \si{10\%} relative error on $n_+^s$, $n_e^s$ and $n_-^s$, \si{20\%} on $\alpha_s$ and \SI{0.2}{V} of RMS error on $V_s$. These errors were then added in quadrature to the uncertainties obtained from the covariance matrix given by the fitting routine.

\section{Application to the experimental data}
\label{sec:data}

\begin{figure}
\centering 
\includegraphics[width=.7\textwidth]{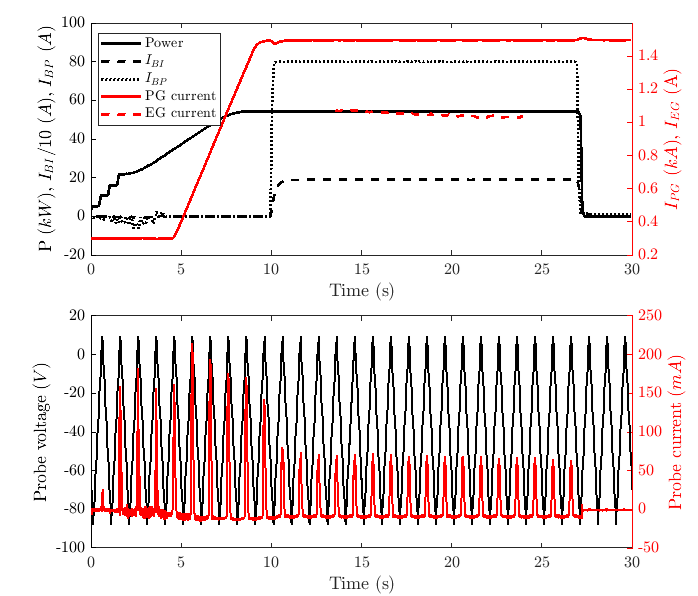}
\caption{\label{fig:plasma_example} Machine parameters (top) and Langmuir probe PG12 signals (bottom) during a plasma blip with $P_{RF}=\SI{50}{kW}/driver$, $I_{BI}=\SI{190}{A}$, $I_{BP}=\SI{80}{A}$, $I_{PG}=\SI{1.5}{kA}$ at \SI{0.3}{Pa} of source filling pressure. Extraction voltage is $V_{EG}=\SI{3}{kV}$.}
\end{figure}

The model was applied to the data collected during SPIDER experimental campaigns, allowing to follow the evolution of plasma parameters during the plasma pulses.
Standard plasma blips of \SI{27}{s} were repeated every 4 minutes during most of the day during the campaigns with cesium evaporation. The plot of machine parameters during a plasma blip is shown in figure~\ref{fig:plasma_example}. The top graph shows the trends of machine parameters: power per driver, plasma grid  bias ($I_{BI}$) and bias plate ($I_{BP}$) current, plasma grid filter current ($I_{PG}$) and extraction grid current ($I_{EG}$), while the bottom graph presents the voltage (in black) and current (in red) signals measured by the Langmuir probe embedded in the center of the plasma grid in correspondence of the top segment.

\begin{figure}
\centering 
\subfloat[\label{fig:fit_example}]{\includegraphics[width=.45\textwidth]{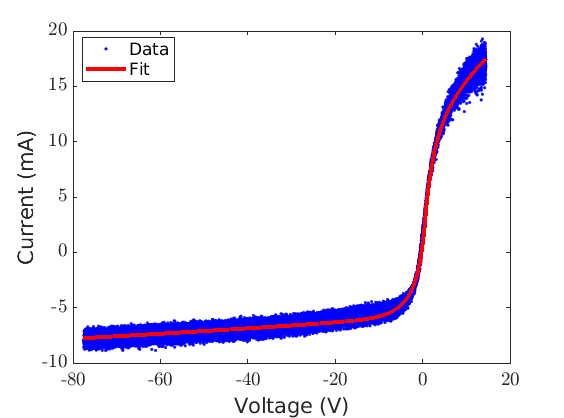}}
\\
\subfloat[\label{fig:trend_blip_density}]{\includegraphics[width=.45\textwidth]{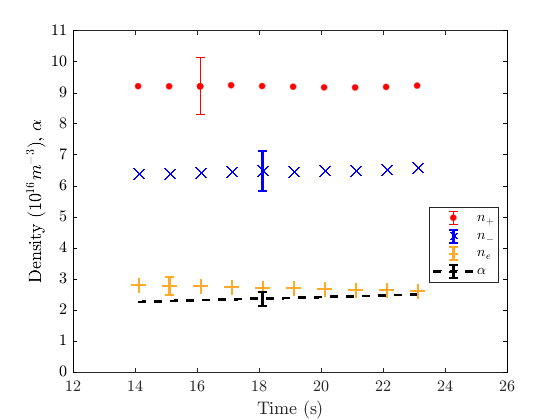}}
\qquad
\subfloat[\label{fig:trend_blip_Te}]{\includegraphics[width=.45\textwidth]{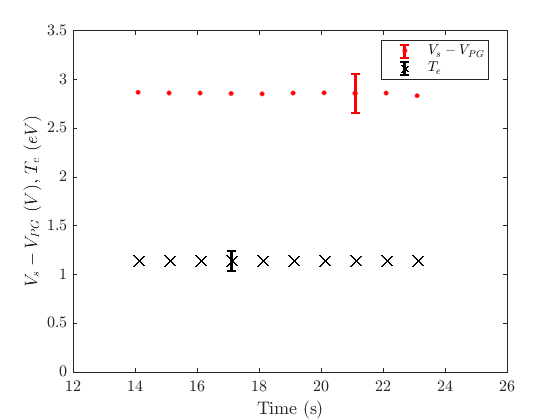}}
\caption{\label{fig:trend_blip} Trend of plasma parameters during the plasma blip of figure~\ref{fig:plasma_example}. (a) Example of the fitted current-voltage characteristic. (b) Densities. (c) Electron temperature, electronegativity and plasma sheath potential referred to the plasma grid.}
\end{figure}

As it can be noticed, from 10 to 28 seconds of each plasma blip the machine parameters were kept constant, and from 14 to 24 seconds beam extraction was performed. This time interval was chosen for the analysis of the probe data. A plot of the fitted plasma parameters for this plasma blip during beam extraction is reported in figure~\ref{fig:trend_blip}, with an example of fitted data reported in figure~\ref{fig:fit_example}. Figure~\ref{fig:trend_blip_density} indicates that positive and negative ion densities remain constant, while the electron density slightly decreases by \SI{9}{\%}. This is in agreement with the trend of the extraction grid current reported in figure~\ref{fig:plasma_example}, which also decreases by \SI{9}{\%}. The estimated electronegativity $\alpha$ is around 2.4, also slightly increasing due to the electron density reduction, while the sheath potential remains at \SI{2.8}{V} with respect to the plasma grid potential (figure~\ref{fig:trend_blip_Te}). The measured electron temperature is \SI{1.2}{eV}, and does not vary during the blip (figure~\ref{fig:trend_blip_Te}).

\begin{figure}
\centering 
\subfloat[\label{fig:std_density}]{\includegraphics[width=.45\textwidth]{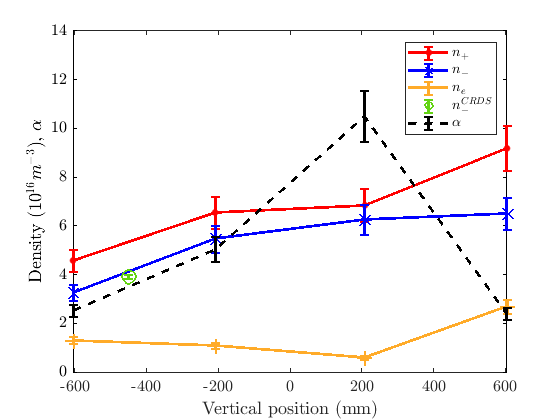}}
\qquad
\subfloat[\label{fig:rev_density}]{\includegraphics[width=.45\textwidth]{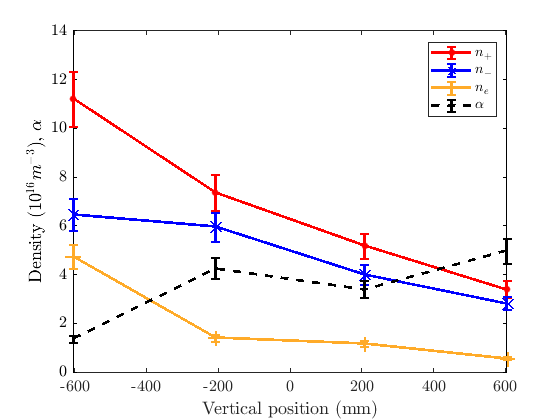}}
\\
\subfloat[\label{fig:std_Te}]{\includegraphics[width=.45\textwidth]{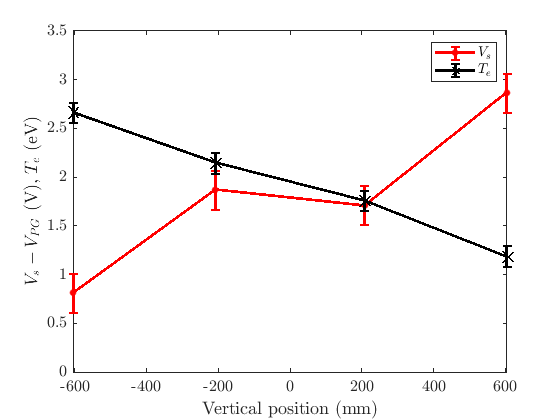}}
\qquad
\subfloat[\label{fig:rev_Te}]{\includegraphics[width=.45\textwidth]{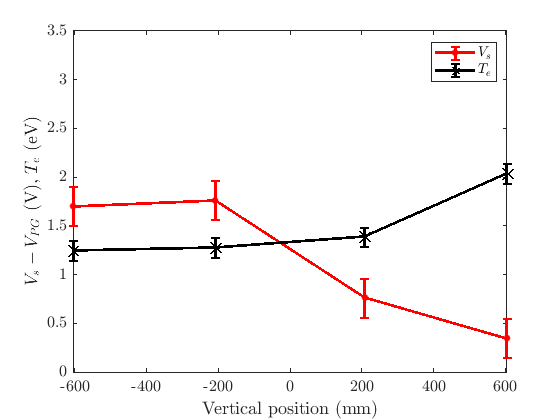}}
\caption{\label{fig:standardVSreversed} Vertical profile of plasma parameters measured for the \textit{standard} (figure a for the densities and electronegativity and c for the electron temperature and sheath potential) and \textit{reversed} (figure b for the densities and electronegativity and d for the electron temperature and sheath potential) configurations. The plasma pulse is performed with \SI{50}{kW/driver} of RF power, $I_{BI}=\SI{190}{A}$, $I_{BP}=\SI{80}{A}$, \SI{0.3}{Pa} of source filling pressure and \SI{1.5}{kA} of filter field current, with opposite direction in the two configurations.}
\end{figure}

A vertical arrangement of four probes in the middle of the plasma grid, one per segment, was used to characterize the vertical plasma uniformity, confirming what was observed also with other diagnostics \cite{art:serianni_2022}. Figure~\ref{fig:standardVSreversed} compares the fit results obtained from these probes for two plasma blips with $P_{RF}=\SI{50}{kW/driver}$, $I_{BI}=\SI{190}{A}$, $I_{BP}=\SI{80}{A}$, same $I_{PG}=\SI{1.5}{kA}$ but opposite direction of the filter field current. Figures~\ref{fig:std_density} and \ref{fig:std_Te} present the plasma parameters obtained for the \textit{standard} operation configuration, while the \textit{reversed} configuration is presented in figures~\ref{fig:rev_density} and \ref{fig:rev_Te}. The standard configuration presents an increase of the densities of positive and negative ions from the bottom to the top. The positive ion density varies from \SI{4.6e16}{m^{-3}} to \SI{9.2e16}{m^{-3}}, while the negative ion density ranges between \SI{3.3e16}{m^{-3}} to \SI{6.5e16}{m^{-3}}, almost doubling its value from top to bottom. This agrees with the single beamlet current measurements reported in~\cite{art:shepherd_2023}: for this source condition the accelerated negative ion current also shows a similar top-bottom asymmetry. The cavity ring-down spectroscopy \cite{art:barbisan_2022} estimation of the negative ion density was also available for this configuration, and its measurement is also reported in figure~\ref{fig:rev_density} (the green point). It provides a measurement of the average negative ion density along a line of sight that crosses horizontally the ion source, at a few millimeters from the plasma grid. Although it is not directly comparable to the probe estimation (it is a measurement of the density inside the bulk plasma, not at the sheath edge, and it is an averaged value at a few millimiters from the plasma grid), the CRDS estimation is of the same magnitude as the Langmuir probe estimation. The potential difference between the sheath edge and the plasma grid is larger in correspondence of the top probe (\SI{2.9}{V}) with respect to the bottom measurement (\SI{0.8}{V}), while it is comparable for the two central segments. The electron temperature varies in the opposite direction, decreasing moving from the bottom segment (\SI{2.6}{eV}) to the top one (\SI{1.2}{eV}). The electron density instead has a minimum in correspondence of the second segment from the top (at \SI{200}{mm}), and correspondingly the electronegativity reaches a value of \si{10.5}.

In the reversed configuration all the trends are the opposite: the positive ion density in the bottom segment (\SI{11.2e16}{m^{-3}}) is larger that that in the top segment (\SI{3.4e16}{m^{-3}}). Negative ion density has a similar trend, but with a smaller excursion (from \SI{6.5e16}{m^{-3}} to \SI{2.8e16}{m^{-3}}), and also the electron density decreases monotonically from the bottom to the top in this configuration (from \SI{4.7e16}{m^{-3}} to \SI{0.5e16}{m^{-3}}). For this reason, the electronegativity does not show the same large variation as for the standard configuration, ranging from 1.8 at the bottom to 4.9 at the top. The trend of the electron temperature and sheath potential is also reversed, with a lower temperature (\SI{1.2}{eV}) and larger potential difference (\SI{1.7}{V}) at the bottom than at the top (\SI{2.0}{eV} and \SI{0.3}{eV}).

\section{Conclusion}
In this paper a fitting procedure for determining the negative ion density from Langmuir probe characteristics was presented, providing a localized measurement of negative ion density. We highlighted the importance of a correct determination of the different collection areas for the various plasma species, especially for the electrons, and applied the model to data collected during SPIDER campaigns. Using the measurement of a vertical array of Langmuir probes, it also allowed to estimate the vertical uniformity of plasma parameters across the plasma grid surface, confirming its dependence on the direction of the filter field current, in agreement with what was found by other diagnostics. The model also showed an agreement between the co-extracted electron current measurement and the electron density measurement of the probe in correspondence of the segment where the plasma density was larger during beam extraction phase. This can be an indication of the use of this analysis technique to assess the uniformity also of the heat load generated by the co-extracted electron current on the extraction grid.

In general, this approach for analyzing Langmuir probe data will provide a new tool to assess the spatial uniformity of negative ion production in a large negative ion source such as SPIDER, using the measurements performed with the bi-dimensional sets of Langmuir probe data embedded in the plasma grid and the bias plate.



\acknowledgments
This work has been carried out within the framework of the ITER-RFX Neutral Beam Testing Facility (NBTF) Agreement and has received funding from the ITER Organization. The views and opinions expressed herein do not necessarily reflect those of the ITER Organization. This work has been carried out within the framework of the EUROfusion Consortium, funded by the European Union via the Euratom Research and Training Programme (Grant Agreement No 101052200 — EUROfusion). Views and opinions expressed are however those of the author(s) only and do not necessarily reflect those of the European Union or the European Commission. Neither the European Union nor the European Commission can be held responsible for them.



\end{document}